\documentclass{elsart}
\usepackage{graphicx}

\begin{document}
\begin{frontmatter}
\title{The unusual temperature dependence of the Eu$^{2+}$   
fluorescence lifetime in CaF$_2$ crystals 
}
\author[cqupt,canterbury]{C.K. Duan}
\author[debye]{A. Meijerink}
\author[canterbury,macdiarmid]{R.J. Reeves}
\author[canterbury,macdiarmid]{M.F. Reid}
\address[cqupt]{Institute of Applied Physics, Chongqing University of Post and Telecommunication, 
Chongqing 400065, P.\ R.\ China}
\address[canterbury]{Department of Physics and Astronomy, University of
Canterbury, Private Bag 4800, Christchurch, New Zealand}
\address[debye]{Debye Institute, Department of Condensed Matter,
Utrecht University, P.O. Box 80,000, 3508 TA Utrecht, The
Netherlands.} 
\address[macdiarmid]{MacDiarmid Institute for Advanced Materials and
Nanotechnology, New Zealand}
\begin{abstract}
  Fluorescence lifetimes of Eu$^{2+}$ doped in CaF$_2$ are measured at
  various temperatures between 4K and 450K.  The lifetime increases
  with between 100K and 300K, in contrast to the usual
  lifetime-temperature dependence. At higher or lower temperatures the
  lifetime decreases with increasing temperature.  The phenomenon is
  explained by simulation of the energy levels and lifetimes of low-lying
  $4f^65d$ states involved in the fluorescence.

\vspace{0.2cm}

{\bf Rare Earth '04 in Nara, Japan, November 7-12, 2004, Paper FP-21\\
Corresponding Author: }\\
Dr.\ Michael F. Reid,  mike.reid@canterbury.ac.nz\\
Department of Physics and Astronomy, University of Canterbury\\
Christchurch, New Zealand.\\
Phone: + 64 3 364 2548  Fax: + 64 3 364 2469

\end{abstract}

\end{frontmatter}

\section{Introduction}

The $4f^N\rightarrow 4f^{N-1}5d$ absorption and $4f^{N-1}5d
\rightarrow 4f^N$ emission spectra of lanthanide ions in crystals are
usually characterized by multi-phonon broad bands as a consequence of
the large difference in the strength of interactions with ligands
between $4f$ and $5d$ orbitals. In most cases the lifetimes of excited
states decrease with increasing temperature as a result of a
substantial increase in non-radiative relaxation rate (the radiative
relaxation rate usually also increases slightly). In contrast,
measurements of Eu$^{2+}$ in CaF$_2$
\cite{Kisliuk1968,Tsuboi1991,Poort1997} showed that the lifetime of
the excited state increases with increasing temperature from about 10K
to room temperature. So far only qualitative explanations have been
given for this phenomenon.

In this work we carry out systematic measurements of excited state
lifetimes of Eu$^{2+}$ doped in CaF$_2$ at various temperatures from
4K to 450K to clarify the experimental results and then calculate of
the energy level structure and lifetime of the $4f^65d$ excited
configuration to simulates the measured results.

\section{Experimental Results}

Laser excitation was provided by a small nitrogen pulsed laser.  Pulse
duration and energy were approximately 2ns and 100 $\mu$J
respectively.  The powdered CaF$_2$:0.1mol\% Eu sample was mounted in
an Oxford Microstat cold finger liquid helium cryostat.  Sample
temperatures could be varied from 5 - 500K using a resistive heater
attached to the cold finger.

The emission from the sample was collected by quartz optics and
detected in two separate systems for fluorescent lifetime or spectral
measurement. A 10 cm Bausch and Lomb monochromator equipped with an
EMI 9659 photomultiplier tube tuned to the peak of the f-d emission
was used for the lifetime measurements.  The PMT signal was averaged
on a digital oscilloscope and downloaded to a computer for processing.

For spectral measurements, the emission was collected by a quartz
fibre optic bundle and analysed by a TRIAX 320 spectrograph, equipped
with a liquid nitrogen cooled CCD.  Measurements of the lifetime and
f-d emission spectrum were recorded at various sample temperatures.

At low temperature ( lower than 50K) the emission spectra are
characterized by sharp zero-phonon line at around 413nm companied by a
broad vibronic band centered at 425nm with some one-phonon lines
clearly distinguishable.  These one-phonon lines and zero-photon line
become indistinguishable around 80K and 120K respectively. The
vibronic band becomes broader as temperature increases.

The decay curves were all found to be exponential with time so each
curve gives a single lifetime for a certain temperature. The lifetime
did not vary across the emission band.  The results are listed in
Table \ref{Lifetime}. It can be seen that between 100K and 320K the
lifetime increases with increasing temperature, which is similar to
early findings\cite{Tsuboi1991}.  However, at temperatures lower than
100K, the measured lifetime shows the usual trend of decreasing with
increasing temperature, different from Ref. \cite{Tsuboi1991}. At
higher temperatures the lifetimes show the usual trend of rapid
decrease with increasing temperature, which we attribute to a large
increase in the nonradiative relaxation rate.

\section{Theoretical simulation and discussion}

\begin{figure}[h]
\caption{
\label{SchematicDiagram}
Schematic diagram of the energy level structure and dynamical
processes for Eu$^{2+}$:CaF$_2$. The ground state is comprised of
three nearly degenerate energy levels of $~^8S_{7/2}$ of $4f^7$ (the
total degeneracy is 8). The energy levels of the excited configuration
$4f^65d$ form a quasi-continuous band starting from around 24,000
cm$^{-1}$.  The conduction band due to ionization of $5d$ electron to
conduction band starts at an energy around 32,000 cm$-1$. The lowest
excited multiplets of $4f^7$ are at around 27,000 cm$-1$, 31,000
cm$-1$ and 34,500 cm$-1$. The excitation forms a thermal equilibrium
distribution among the lowest excited $4f^65d$ energy levels due to
fast non-radiative relaxation among them. These states either relax to
the ground states by emitting a photon or ionize to the conduction
band.}  
\centerline{\includegraphics[width=16cm]{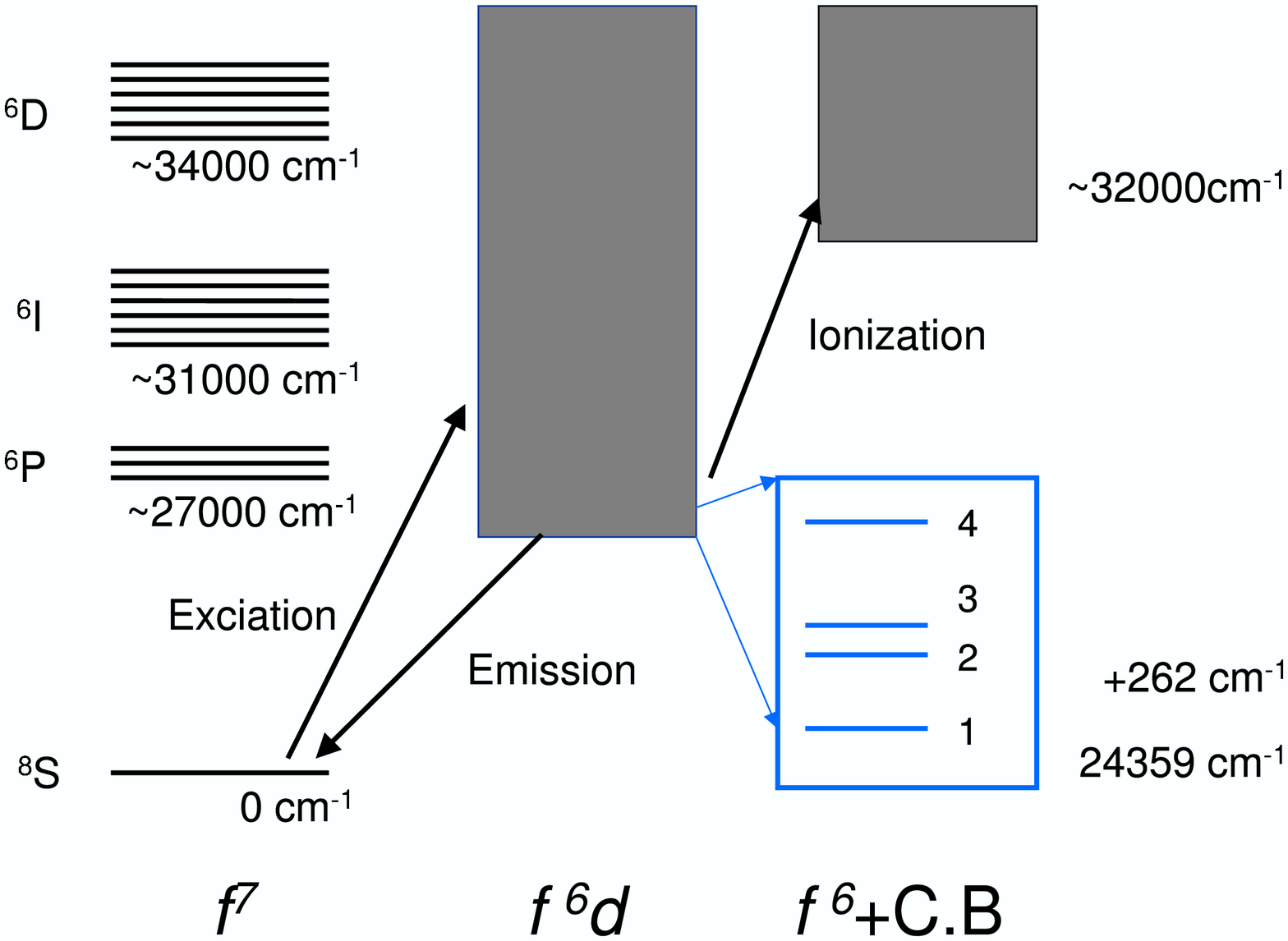}}
\end{figure}

The dynamical processes of excitation in Eu$^{2+}$:CaF$_2$ are show in
Fig. \ref{SchematicDiagram}. The lowest excited states belong to the
$4f^65d$ configuration. The excitation to $4f^65d$ bands quickly relax
to the bottom of the band and than relax radiatively to the ground
states of $4f^7$ configuration by emitting a photon or nonradiatively
by either a multi-phonon processes or by ionization of the $5d$
electron to conduction band.\cite{Dujardin1993,Fuller1987}

Methods for simulating one-photon and two-photon absorption spectra of
Eu$^{2+}$ in CaF$_2$ have been developed using an extended
crystal-field
Hamiltonian\cite{Reid2000,Pieterson2002a,Pieterson2002b,Reid2002}
which takes into account major interactions in both $4f^N$ and
$4f^{N-1}5d$ configurations and the interaction between them
\cite{Burdick2003}. The simulations have successfully explained energy
level positions, one-photon and two-photon transition intensities and
line widths.\cite{Burdick2003}

There are around 30,000 states in $4f^65d$ configuration. The lowest
few hundred of them in the range of 24,000 cm$^{-1}$--30,000 cm$^{-1}$
are spin octet states and have a large line strength for transitions
from and to the near-degenerate ground octet multiplet $~^8S_{7/2}$ of
$4f^7$ configuration.\cite{Dujardin1993} Because of fast nonradiative
relaxation among these $4f^65d$ states, there is thermal equilibrium
among these $4f^65d$ excited states before fluorescence occurs. This
explains why the fluorescence lifetime does not vary across the
emission band.  Radiative relaxation from each of these excited states
gives a broad band, with zero-phonon and some one-phonon lines
distinguishable at only low temperatures.  A superposition of the
transitions allows us to calculate the emission spectrum and lifetime
as a function of temperature.  The lowest few $4f^65d$ energy levels,
their degeneracy, and relative radiative relaxation rates are
presented in Table \ref{EnergyLevels}.  The relaxation rate of the
thermal-equilibrium states can be written as
\begin{equation}
R_{R}= \frac{1}{\tau_1} \frac{\sum\limits_i g_iR_i \exp(-\Delta E_i/kT)}
                {\sum\limits_i g_i \exp(-\Delta E_i/kT)}.
\end{equation}
In this expression $\tau_1$ is the lifetime at 0 K, $g_i$ is the
degeneracy of the state, $R_i$ the relative radiative decay rate,
$\Delta E_i$ the energy difference between the $i$th state and the lowest
excited state, $k$ Boltzmann's constant, and $T$ the temperature.

The dramatic decrease of lifetimes with increasing temperature above
room temperature is common in fluorescence materials. It is usually
due to a rapid increase in nonradiative relaxation rate.  As the
temperature increases, both the nonradiative multiphonon relaxation
rate\cite{Sturge1973} and the nonradiative relaxation of excitation
due to ionization of $5d$ electron to conduction band
\cite{Fuller1987} increase sharply at high temperature.
Phenomenologically, the nonradiative relaxation rate $R_{NR}$ can be
written as:
\begin{equation}
R_{NR}=W_{NR}\exp (-E_I/kT),
\end{equation}
where the non-radiative rate constant $W_{NR}$ and the activation energy $E_I$
are treated as parameters. This mechanism only contributes significantly at
room temperature or even higher temperature.

The electronic energy levels also couple with vibrations. At zeroth
order approximation, the electronic and vibrational states can be
considered as two independent systems and the lifetime of the states
of the electron-phonon (EP) system is independent of the vibronic
states.  However, at higher-order approximations, especially when the
electronic states and the vibrational modes are degenerate, the
coupled EP states split into several states, the first few of which
separate from each other by a few to a hundred
wavenumbers.\cite{Chase1970} These coupled EP states can have slightly
different radiative relaxation rates. This effect may be modelled
phenomenologically as follows:
\begin{equation}
\Delta R = \frac{1}{\tau_1} A [1-\exp(-\Delta E_A/kT)],
\end{equation}
where, again, $A$ and $E_A$ are treated as fitting parameters. This
mechanism contributes significantly only at low temperature.

The lifetime at any temperature can be calculated by combining the
above rates to give
\begin{equation}
\tau(T) = \frac{1}{R_R(T) + \Delta R (T) +R_{NR}(T)},
\end{equation}
with five fitting parameters $\tau_1$, $W_{NR}$, $E_I$, $A$ and $E_A$. 

\begin{figure}[h]
\caption{ Measured and calculated lifetimes as a function of temperature. Measured lifetimes
  are shown as circles. The calculated lifetimes are shown as a curve.
  The values of parameters are $\tau_1= 644$ ns, $A=-0.05$, $E_A=20
  {\rm cm}^{-1}$, $W_{\rm NR} =1.23 \times 10^{14} {\rm s}^{-1}$ and $ E_I
  = 4370 {\rm cm}^{-1}$
\label{LifetimeFigure}
}
\centerline{\includegraphics[width=16cm]{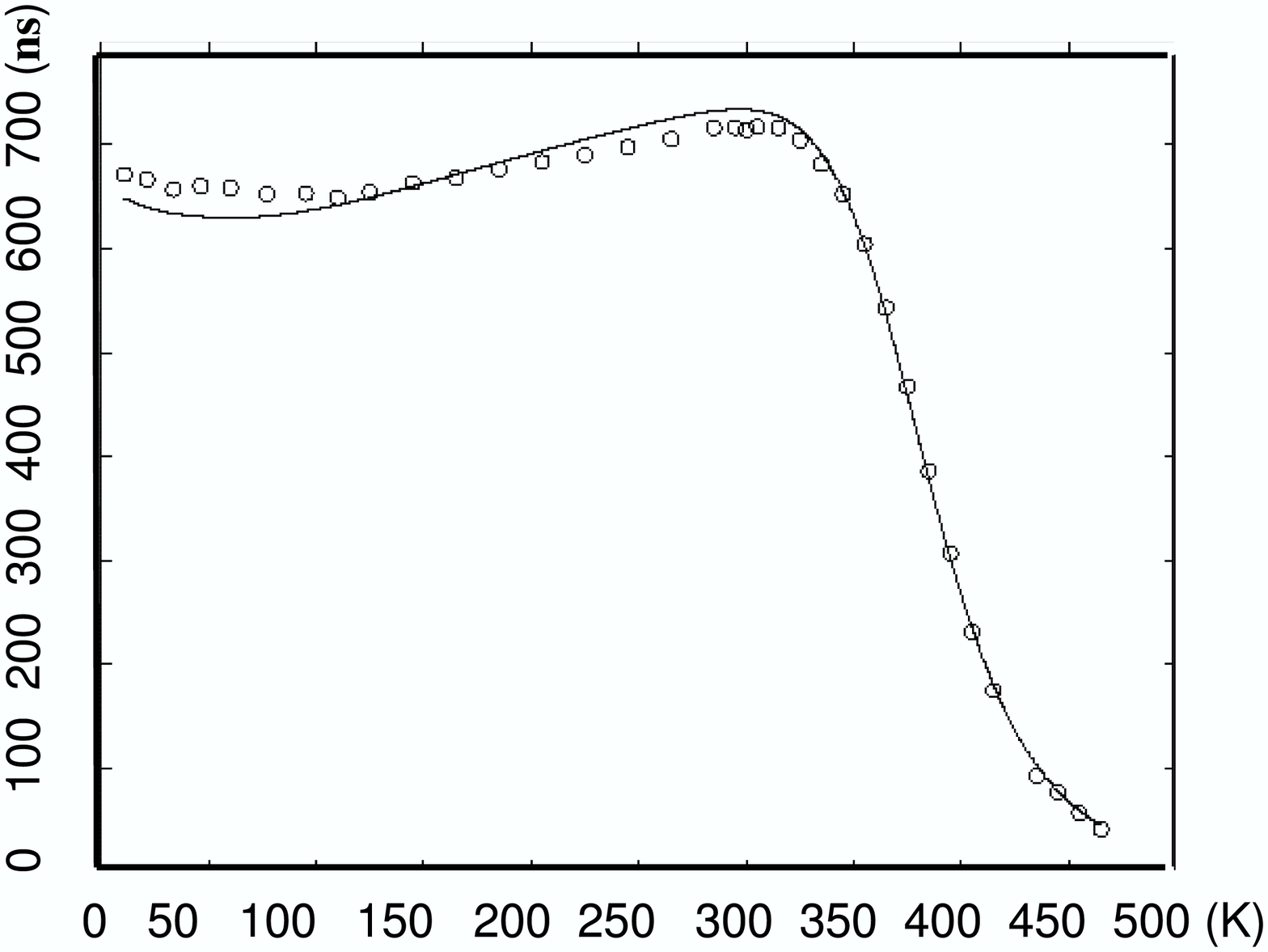}}
\end{figure}

The experimental results and the fitted results are plotted in Fig.
\ref{LifetimeFigure}. The parameter values used are listed in the
figure caption.  It can be seen that the calculated curve gives the
right trend of lifetime as a function of temperature. However, the
calculated lifetime increase is greater than the measured lifetime. A
possible explanation is that the radiative transition rate of each
state increases due to a deviation from the Condon approximation.

\section{Conclusion}

The fluorescence lifetime of the $4f^65d\rightarrow 4f^7$ transition
of Eu$^{2+}$ doped in CaF$_2$ has been measured at various temperatures
between 4K and 450K. Between 100K and 300K the lifetime increases with
increasing temperature, different from the usual lifetime-temperature
dependence. Energy levels of $4f^7$ and $4f^65d$ configurations and
transitions rates between them have been calculated using an extended
crystal-field Hamiltonian.  The results show that several excited
$4f^65d$ energy levels separated by energies of a few hundred
wavenumbers contribute to the excited state lifetimes.  The lowest
excited state has a much larger radiative decay rate than the other
excited states that are thermally populated.  As temperature
increases, the chance of initial states occupying higher excited
energy levels, which have a smaller radiative decay rate, increases
and therefore the average radiative decay rate decreases. Between 100K
and 320K in CaF$_2$, the radiative decay dominates the $5d\rightarrow
4f$ excitation relaxation and therefore we observe an increasing
lifetime with increasing temperatur.  At low or high temperatures the
lifetime decreases with increasing temperature due to thermal
activation and quenching effects.

\newpage

\newpage

\newpage 
\begin{table}[ht]
\caption{
The lifetime of Eu$^{2+}$:CaF$_2$ measured at different temperatures:
$T$: temperature; $\tau (T)$: lifetime at $T$.
\label{Lifetime}}
\begin{tabular}{cc|cc|cc}
\hline
$T$ (K) & $\tau$ (ns) &$T$ (K)  &$\tau$ (ns) &$T$ (K)  &$\tau$ (ns)\\
\hline
  10.8 & 671 & 21.2 & 666  &33.6  & 657\\
  46.3 & 660 & 60.1 & 658  &77.1  & 652\\ 
  95.2&  653 & 110  &  649 & 125  & 654\\
  145 &  663 & 165  &  668 & 185 &  676\\
  205 &  683 &  225 &  690 & 245 &  697\\
  265&   705 &  285 &  716 &  295&  716\\
  300 &  714 &  305 &  717 &  315 & 716\\
  325 &  704 &  335 &  681&  345 &  652\\
  355 &  604 &  365 &   543&  375 & 467\\
  385 &  386 &  395 &  307&  405&   231\\
  415 &  175 &   435&   93 &  445 &  77 \\
  455 &  57  &  465 &  42\\
\hline
\end{tabular}
\end{table}
\newpage
\begin{table}[ht]
\caption{\label{EnergyLevels}
Calculated energy levels, their relative radiative relaxation rates 
 and degeneracies, where $i$, $g_i$, $E_i$ $\Delta E_i$ and $R_i$
are energy level No., degeneracy, energies, energies relative to 
the first energy level and relative relaxation rate respectively.
The units of $E_i$ and $\Delta E_i$ are $cm^{-1}$. 
Only the lowest few energy levels contribute to the lifetime. Note: the absolute
relaxation rate is proportional to the square of the 
radial integral $\langle f | r | d \rangle$,  which cannot be
 calculated accurately.
It is treated as an adjustable parameters in this work.
} 
\begin{tabular}{rrrrr}\hline
$i$ & $g_i$ & $E_i$ & $\Delta E_i$ & $R_i$\\
\hline
1     &   4     &    24359    &          0     &        1.000\\
2     &   4     &    24621    &        262     &       0.128\\
3     &   2     &    24668    &        309     &       1.528\\
4     &   2     &    25007    &        648     &       0.496\\
5     &   4     &    25046    &        687     &        0.472\\
6     &   2     &    25387    &       1028     &        1.043\\
\hline
\end{tabular}
\end{table}


\begin{thebibliography}{10}
\bibitem{Kisliuk1968}
   P. Kisliuk et. al, Phys. Rev. 171, 336 (1968)
\bibitem{Tsuboi1991}
   T. Tsuboi and P. Silfsten, J. Phys. Condens. Matter 3, 9163 (1991).
\bibitem{Reid2000}M.\ F.\ Reid, L. van Pieterson, R.\ T.\ Wegh, A.\ Meijerink, Phys.\ Rev.\
                  B 62, 1291 (2000).
\bibitem{Poort1997} S.H. Poort, A. Meijerink and G. Blasse, J. Phys. Chem Solids, 58, 1451 (1997).
\bibitem{Pieterson2002a}L. van Pieterson, M.F. Reid, R.T. Wegh, S. Soverna,
                  A. Meijerink, Phys. Rev. B 65, 045113 (2002).
\bibitem{Pieterson2002b}L. van Pieterson, M.F. Reid, G.W. Burdick, A. Meijerink, Phys. Rev. B 65, 045114 (2002).
\bibitem{Reid2002}M.F. Reid, L. van Pieterson, A. Meijerink, J. Alloys Compounds 344, 240 (2002).
\bibitem{Burdick2003} G.W. Burdick, A. Burdick, C.K. Duan, and M.F. Reid, {\it Many-body perturbation theory for spin-forbidden two-photon
       spectroscopy of $f$-element compounds and its application to
       Eu$^{2+}$ in CaF$_2$ } (unpublished).
\bibitem{Dujardin1993} C. Dujardin, B. Moine and C. Pedrini, J. Lumin. 54, 259 (1993).
\bibitem{Fuller1987} R.\ Fuller and D.\ S.\ McClure, J.\ Luminescence {38}, 193 (1987)
\bibitem{Sturge1973} M.D. Sturge, Phys. Rev. B 8, 6 (1973).
\bibitem{Chase1970} L.\ L.\ Chase,  Phys. Rev. B 2, 2308 (1970). 
\end{thebibliography}
\end{document}